\documentclass[a4paper]{PoS}

\usepackage{xspace}
\usepackage{amsmath}
\newcommand{\mumu}{$\mu^+\mu^-$\xspace}
\newcommand{\ee}{$\mathrm{e^+e^-}$\xspace}
\newcommand{\emu}{$\mathrm{e^{\mp}}\mu^{\pm}$\xspace}

\newcommand{\pb}{\mbox{\ensuremath{\,\text{pb}}}\xspace}
\newcommand{\TeV}{\mbox{\ensuremath{\,\text{TeV}}}\xspace}
\newcommand{\GeV}{\mbox{\ensuremath{\,\text{GeV}}}\xspace}

\newcommand{\stat}{\mbox{\ensuremath{\,\text{(stat)}}}\xspace}
\newcommand{\syst}{\mbox{\ensuremath{\,\text{(syst)}}}\xspace}
\newcommand{\lum}{\mbox{\ensuremath{\,\text{(lum)}}}\xspace}
\newcommand{\ttbar}{\ensuremath{\rm t\bar{t}}\xspace}

\newcommand{\resultxsectopmass}{\ensuremath{815 \pm  2 \stat \pm 29 \syst \pm 20 \lum \pb}\xspace}
\newcommand{\resulttopmassMC}{\ensuremath{172.33 \pm  0.14 \stat \pm^{0.66}_{0.72} \syst \GeV}\xspace}

\newcommand{\as}{\ensuremath{\alpha_\mathrm{S}}\xspace}

\newcommand{\asmz}{\ensuremath{\alpha_\mathrm{S}(m_\mathrm{Z})}\xspace}

\newcommand{\stt}{\ensuremath{\sigma_\mathrm{t\bar{t}}}\xspace}

\newcommand{\Mtt}{\ensuremath{M(\ttbar)}\xspace}
\newcommand{\ytt}{\ensuremath{y(\ttbar)}\xspace}
\newcommand{\Njet}{\ensuremath{\mathrm{N}_\text{jet}}\xspace}

\newcommand{\pt}{\ensuremath{\mathrm{p_T}}\xspace}

\newcommand{\msbar}{\ensuremath{\mathrm{\overline{MS}}}\xspace}
\newcommand{\mtmt}{\ensuremath{m_\mathrm{t}(m_\mathrm{t})}\xspace}

\newcommand{\mtp}{\ensuremath{m_\mathrm{t}^{\mathrm{pole}}}\xspace}
\newcommand{\mtmc}{\ensuremath{m_\mathrm{t}^{\mathrm{MC}}}\xspace}

\newcommand{\mt}{\ensuremath{m_\mathrm{t}}\xspace}

\newcommand{\mlb}{\ensuremath{m_\mathrm{\ell b}^{\mathrm{min}}}\xspace}

\title{Probing QCD using top quark pair production at $\bf \sqrt{s}=13~\mathrm{\textbf{TeV}}$ in CMS}

\ShortTitle{Short Title for header}

\author{\speaker{Matteo M. Defranchis} \\
        Deutsches Elektronen-Synchrotron (DESY) \\ 
         Notkestra{\ss}e 85, 22607 Hamburg - Germany\\
        E-mail: \email{matteo.defranchis@cern.ch}\\
        on behalf of the CMS Collaboration
        
        }

\abstract{
	Measurements of the top quark-antiquark pair production cross section, \stt, can be used to constrain the strong coupling constant, \as, the top quark mass, \mt, and the parton distribution functions (PDFs). In this poster, the two most recent relevant results published by the CMS Collaboration are presented. The analyses are performed using proton-proton collision data at a centre-of-mass energy of 13\TeV recorded by the CMS detector at the CERN LHC in 2016, corresponding to an integrated luminosity of $35.9~\mathrm{fb^{-1}}$. In the first one,  \as and \mt are extracted independently from a measurement of the inclusive \stt, using next-to-next-to-leading order theoretical predictions. In the second, a measurement of the normalized triple-differential \ttbar cross section is performed; the result is then used together with HERA deep-inelastic scattering data to perform a simultaneous determination of \as, \mt, and the PDFs, at next-to-leading order. As a result, the uncertainty in the gluon PDF and its correlation with \as are significantly reduced at high parton momentum fraction, the kinematic range probed by \ttbar production. The result also yields the most precise determination of the top quark pole mass, to date.

	}

\FullConference{
European Physical Society Conference on High Energy Physics - EPS-HEP2019 -\\
			10-17 July, 2019\\
			Ghent, Belgium}

\begin{document}

\section{Introduction}

The production of top quark-antiquark (\ttbar) pairs in proton proton (pp) collisions can be calculated in perturbative quantum chromodynamics (pQCD), and depends on the top quark mass, \mt, the strong coupling constant, \as, and the parton distribution functions (PDFs) of the proton. Calculations of inclusive and differential cross sections are available up to next-to-next-to-leading order (NNLO) in perturbation theoretical, and fits of theory predictions to measured observables can be used to extract the QCD parameters (\as, \mt) and the proton PDFs. The two most recent results by the CMS Collaboration obtained using pp collision data at the centre-of-mass energy of 13~TeV are presented in this poster. In the first one, described in Section~\ref{sec:incl}, \mt and \as are extracted independently from a measurement of the inclusive \ttbar production cross section, \stt, using different sets of PDFs. In the second, described in Section~\ref{sec:diff}, a measurement of the normalized triple-differential \ttbar cross section is used together with HERA deep-inelastic scattering data to perform a simultaneous determination of \as, \mt, and the proton PDFs. 

\section{Extraction of \as and \mt from the inclusive \stt}
\label{sec:incl}

In this analysis, described in details in Ref.~\cite{Sirunyan:2018goh}, \stt is measured simultaneously with the top quark mass in the simulation, \mtmc, via a likelihood fit to multi-differential distributions of final state observables. The measurement is performed using pp collisions data recorded by the CMS detector~\cite{Chatrchyan:2008aa} in 2016, corresponding to an integrated luminosity of $35.9~\mathrm{fb^{-1}}$. Candidates of \ttbar events are selected in the the final state containing an electron and a muon of opposite charge (\emu). The fit is  performed in categories of jet and b~tagged jet multiplicities, and jet \pt spectra are used to constrain the jet energy scale. Furthermore, the \mlb distribution, i.e. the minimum invariant mass found when combining a b~tagged jet and a lepton, is used to measure \mtmc. Systematic uncertainties are constrained within the visible phase space, while additional uncertainties are assigned to the extrapolation to the full phase space. This method allows to measure \stt at the optimal value of \mtmc, thus reducing the experimental dependence on this parameter. The fit yields:
\begin{align*}                     
	\stt  & =  \resultxsectopmass, \\                                                                                                                                       
	\mtmc & =  \resulttopmassMC,                                                                                                                                           
\end{align*}
where the uncertainties refer to the statistical (stat), the systematic (syst), and the luminosity (lum) uncertainties.

\begin{figure}[h!]
	\includegraphics[width=.5\textwidth]{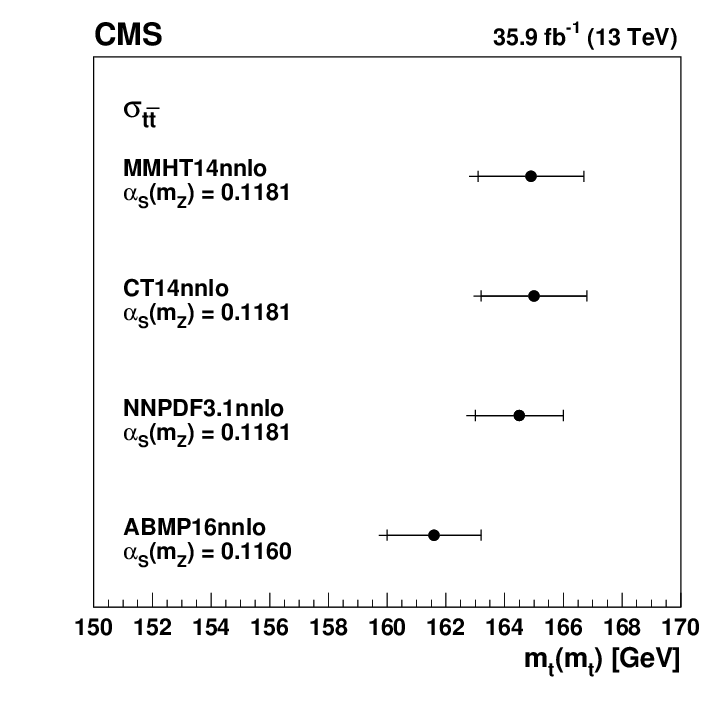}
	\includegraphics[width=.5\textwidth]{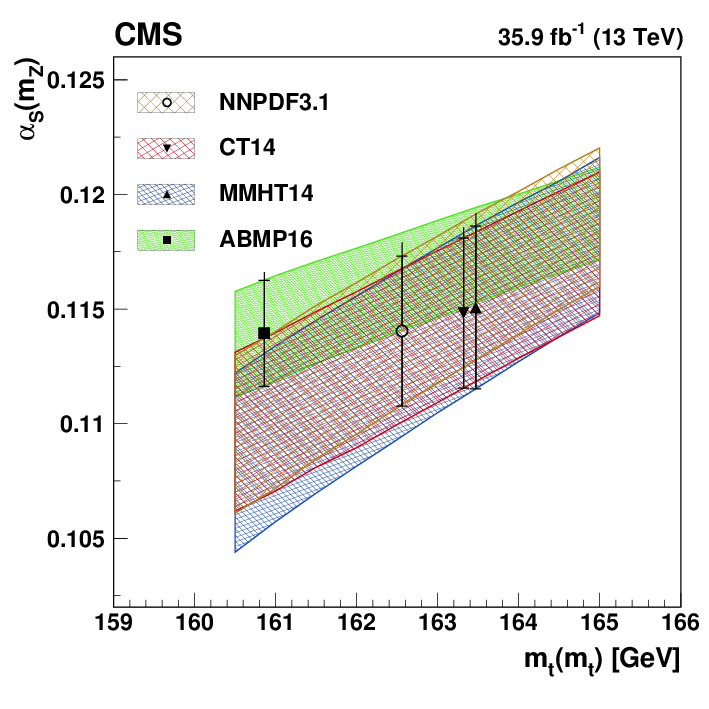}
	\caption{Extracted values of \mtmt (left) and \asmz (right), obtained using calculations in the \msbar scheme with different sets of PDFs. The dependence of the extracted \asmz on the assumed \mtmt is represented by the shaded bands~\cite{Sirunyan:2018goh}.
		\label{fig:as_mt}}
\end{figure}

The measured \stt is then used to extract \as and \mt by comparing to NNLO calculations performed in the modified minimal subtraction (\msbar) scheme with different sets of PDFs. Since it is not possible to simultaneously extract \as and \mt from \stt, one of the two parameters is fixed to the value used in the PDFs when the other one is extracted. The resulting values of \mtmt are shown in Figure~\ref{fig:as_mt} (left). 
The best-fit values of \as at the Z~boson mass, \asmz, correspond to the marker in Figure~\ref{fig:as_mt} (right), while the shaded bands represent the dependence of the extracted \as on the assumed value of \mtmt.
These results are the most precise determination of \mtmt, to date, and the most precise NNLO extraction of \asmz at a hadron collider.
The extraction of \mt was also performed in the on-shell scheme, in order to extract the top quark pole mass. The results can be found in Ref.~\cite{Sirunyan:2018goh}. 

\section{Simultaneous determination of \as, \mt, and the proton PDFs}
\label{sec:diff}

\begin{figure}[h!]
\centering	\includegraphics[width=\textwidth]{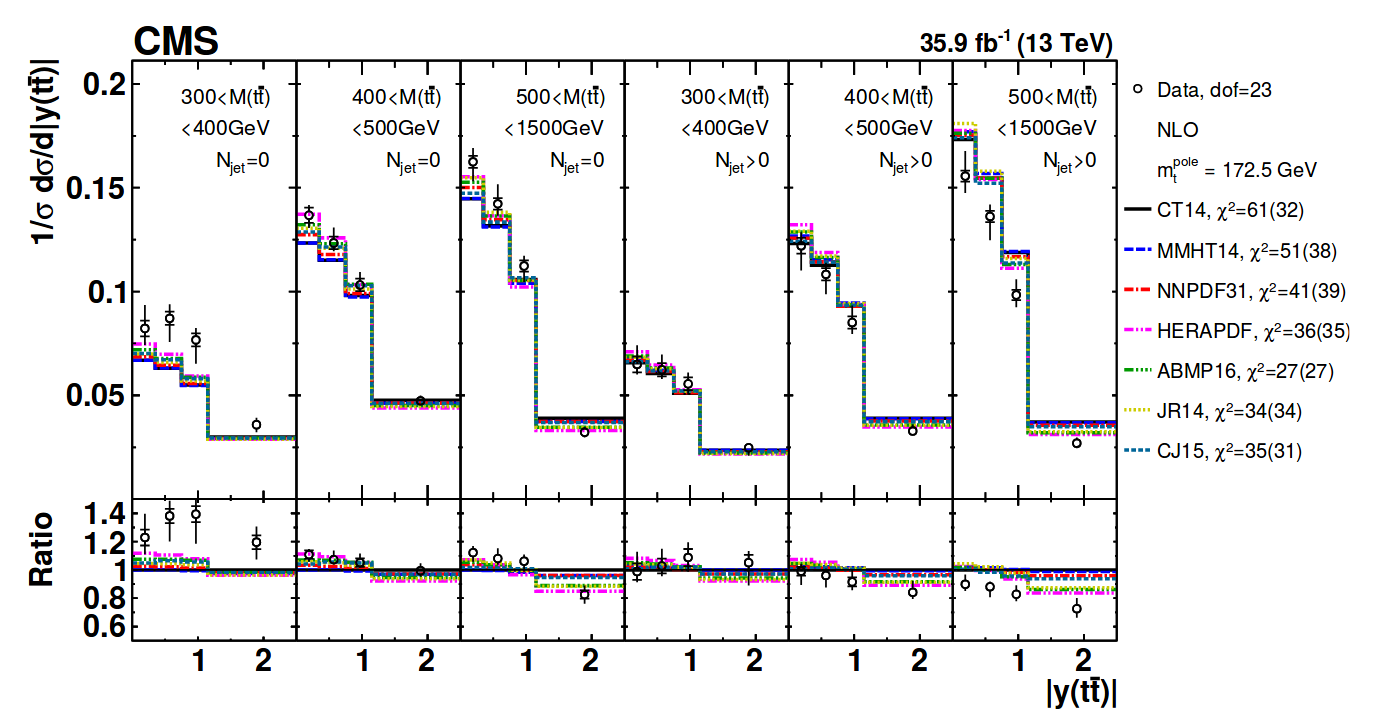}
\caption{Normalized triple-differential \ttbar cross section as a function of \Mtt, \ytt, and \Njet, unfolded to the parton level. The result is compared to NLO theoretical predictions obtained with different sets of PDFs~\cite{Sirunyan:2019zvx}.
	\label{fig:diff_xsec}}
\end{figure}

In the analysis described in detail in Ref.~\cite{Sirunyan:2019zvx}, \as, \mt, and the proton PDFs are simultaneously extracted from a measurement of the normalized triple-differential \ttbar cross section as a function of the invariant mass of the \ttbar system, \Mtt, the rapidity of the \ttbar system, \ytt, and the multiplicity of jets not originating from the decays of the top quarks, \Njet. In particular, the threshold of the \Mtt distribution is sensitive to \mt, \Njet strongly depends on \as, and a combination of \Mtt and \ytt is sensitive to the parton momentum fraction $x$~\cite{Sirunyan:2019zvx}. The measurement is performed using the same data set as the analysis described in Section~\ref{sec:incl}, selecting \ttbar candidate events in the \emu, \ee, and \mumu channels. In Figure~\ref{fig:diff_xsec}, the measured cross section unfolded to the parton level is compared to theoretical predictions at next-to-leading order (NLO) obtained with different sets of PDFs.

The QCD parameters and the proton PDFs are then determined simultaneously via a  fit of NLO theoretical predictions in the on-shell scheme to the measured differential cross section and HERA deep-inelastic scattering (DIS) data. The HERA data are sensitive to the PDFs of the quarks, while \ttbar data provide additional information about the PDF of the gluon, which cannot be probed directly in DIS. The QCD parameters are determined with remarkable precision:
\begin{align*}
\asmz =& ~0.1135{}~^{+0.0021}_{-0.0017},\\
\mtp =& ~170.5 \pm 0.8 \GeV,
\label{eq:as-nom}
\end{align*}
where \mtp denotes the top quark pole mass. The largest contributions to the total uncertainty arise from the systematic uncertainty in the measured cross section and the variation of the factorization and renormalization scales in the NLO calculation. The fit yields the most precise determination of \mtp, to date.

\begin{figure}[h!]
	\includegraphics[width=.5\textwidth]{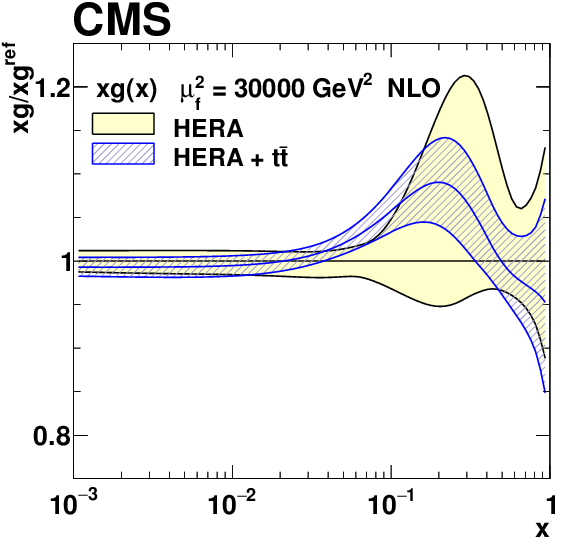}
	\includegraphics[width=.5\textwidth]{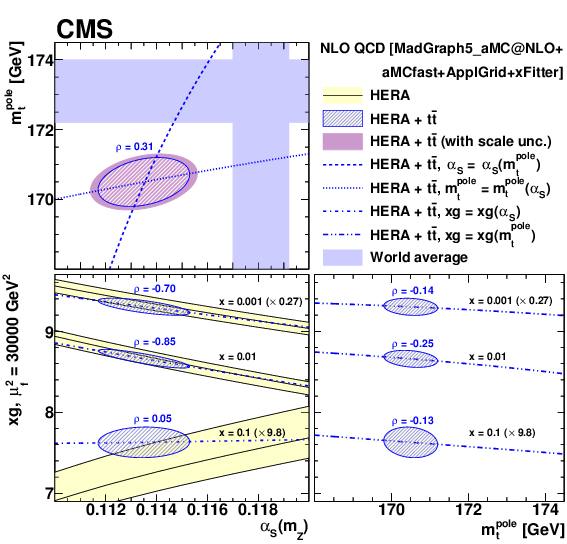}
	\caption{Gluon PDF and its uncertainty as obtained from the combined fit to HERA \ttbar data, compared to the one obtained using HERA data only (left). Correlation between \as, \mt, and the gluon PDF after the fit, with and without including \ttbar data (right)~\cite{Sirunyan:2019zvx}. 
		\label{fig:PDF}}
\end{figure}

In Figure~\ref{fig:PDF} (left), the gluon PDF obtained in the combined fit to HERA and \ttbar data is compared to the result obtained considering HERA data only. Significant improvement is achieved around $x = 0.1$, which corresponds to the kinematic region of \ttbar production at 13~TeV.
In the same range, the correlation between the gluon PDF and \as is significantly reduced, as shown in Figure~\ref{fig:PDF} (right). The correlations between \as and \mtp, and between \mtp and the gluon PDF are also shown.

\section{Summary}

This poster illustrates the two most recent results published by the CMS Collaboration in which top quark-antiquark (\ttbar) data are used to probe quantum chromodynamics. The measurements are performed using proton-proton collision data at $\sqrt{s} =  13~\mathrm{TeV}$ recorded by the CMS detector at the CERN LHC in 2016, corresponding to an integrated luminosity of $35.9~\mathrm{fb^{-1}}$. These analyses resulted in the most precise measurements of the top quark mass in the \msbar and on-shell schemes, to date, and in the most precise measurement of the strong coupling constant, \as, at a hadron collider. Furthermore, through the simultaneous determination of \as, the top quark mass, and the proton distribution function (PDFs), the uncertainty in the gluon PDF and its correlation with \as are significantly reduced in the relevant kinematic range.


\begin{thebibliography}{99}
\bibitem{Chatrchyan:2008aa}
CMS Collaboration,
``The CMS Experiment at the CERN LHC'',
JINST {\bf 3} (2008) S08004,
doi:10.1088/1748-0221/3/08/S08004 .

\bibitem{Sirunyan:2018goh}
CMS Collaboration,
``Measurement of the $\mathrm{t}\overline{\mathrm{t}}$ production cross section, the top quark mass, and the strong coupling constant using dilepton events in pp collisions at $\sqrt{s} =$ 13 TeV'',
Eur.\ Phys.\ J.\ C {\bf 79} (2019) 368,
doi:10.1140/epjc/s10052-019-6863-8
[{\tt arXiv:1812.10505}].

\bibitem{Sirunyan:2019zvx}
CMS Collaboration,
``Measurement of $\mathrm{t\bar t}$ normalised multi-differential cross sections in pp collisions at $\sqrt s=13$ TeV, and simultaneous determination of the strong coupling strength, top quark pole mass, and parton distribution functions'',
Submitted to: Eur. Phys. J.
[{\tt arXiv:1904.05237}].


\end{thebibliography}
\end{document}